%% file: paper.tex
\documentclass[11pt]{article}
\usepackage[margin=2cm]{geometry}
\usepackage{microtype}
\usepackage{cite}
\usepackage{amsmath,amssymb,amsfonts}
\usepackage{algorithmic}
\usepackage{graphicx}
\usepackage{textcomp}
\usepackage{xcolor}
\usepackage{booktabs}
\usepackage{multirow}
\usepackage{array}
\usepackage{url}
\usepackage{hyperref}
\usepackage{tikz}
\usetikzlibrary{shapes.geometric, arrows.meta, positioning, fit, calc, backgrounds}
\usepackage{amsthm}
\newtheorem{definition}{Definition}
\newtheorem{proposition}{Proposition}

\usepackage{authblk}

\hypersetup{
    colorlinks=true,
    linkcolor=blue,
    citecolor=blue,
    urlcolor=blue
}

\begin{document}

\title{\textbf{The Accessibility Capability Boundary:\\
Operational Limits and Expansion Potential of\\
AI-Generated Browser-Native Accessibility Systems}}

\author[1]{Rizwan Jahangir}
\author[2]{Daisuke Ishii}

\affil[1]{NUST Business School, NUST, Islamabad, Pakistan}
\affil[2]{Kiara Inc., Tokyo, Japan}

\maketitle

\begin{abstract}
As large language models (LLMs) demonstrate increasing competence in synthesizing functional user interfaces, a fundamental question emerges in accessibility computing: \textit{how far can AI-driven accessibility systems go?} This paper introduces the \textit{Accessibility Capability Boundary} (ACB), a formal framework for reasoning about the operational limits and expansion potential of autonomous accessibility systems, and grounds this theory in a real-world systems artifact. We model accessibility not as a binary compliance property but as a dynamic, multidimensional capability space constrained by measurable variables including deployment latency, cognitive load, infrastructure dependency, offline persistence, interaction complexity, and adaptability. We argue that AI-generated, browser-native systems---constructed as single-file HTML artifacts leveraging standard browser APIs---may dramatically shift the ACB outward by reducing deployment friction to near-zero and enabling rapid, context-specific interface adaptation. We ground our theoretical framework in the analysis of two real-world exploratory prototypes. The first is an AI-generated browser-native accessibility interface deployed for a blind user in Nepal. The second is a fully functional, open-source webcam alignment assistant for visually impaired users, serving as a concrete systems artifact. Through formal definitions, propositions, and a comparative evaluation matrix, we characterize the regions of the accessibility capability space that such systems can and cannot reach. We further identify remaining computational, infrastructural, and verification constraints that constitute the hard boundaries of this paradigm. This work contributes a theoretical foundation for understanding the scalable limits of autonomous accessibility computing and proposes a research agenda for future work in accessibility-aware AI systems.
\end{abstract}

\vspace{1em}
\noindent\textbf{Keywords:} accessibility computing, artificial intelligence, browser-native systems, capability boundaries, large language models, user interfaces, assistive technology, offline-first, progressive web apps, HCI

\section{Introduction}
\label{sec:intro}

The design and deployment of accessible digital systems has historically been a labor-intensive, compliance-driven engineering effort. The dominant paradigm requires developers to manually implement accessibility standards---most prominently the Web Content Accessibility Guidelines (WCAG) \cite{Caldwell2008} and the Accessible Rich Internet Applications specification (WAI-ARIA) \cite{W3C_ARIA}---and to test resulting systems against a fragmented landscape of assistive technologies including screen readers, switch access devices, and braille displays \cite{Trewin2013}. Despite decades of research and standardization effort, the web and digital ecosystem at large remain substantially inaccessible to users with disabilities \cite{Lazar2007, Bigham2011}.

The emergence of large language models (LLMs) capable of synthesizing complex, functional HTML, CSS, and JavaScript from natural language prompts introduces a qualitatively different capability: the autonomous generation of accessible systems on demand \cite{Jiang2026, Calo2023}. An interface that previously required weeks of development effort can now be prototyped in minutes. This compression of the development cycle has profound implications for accessibility, particularly in low-resource contexts where professional developers with accessibility expertise are scarce \cite{Pal2016, Karki2023}.

This paper investigates the central question: \textit{Can we determine practical, theoretical, computational, and infrastructural bounds for autonomous accessibility systems generated by AI?} Rather than claiming that AI has solved accessibility, we propose a formal framework for reasoning about what such systems can and cannot achieve. We introduce the \textit{Accessibility Capability Boundary} (ACB), a concept that models accessibility as a dynamic operational space with measurable constraints and expansion potential.

Our investigation is grounded in the analysis of two real-world exploratory prototypes. The first is a browser-native accessibility interface generated using Claude (Anthropic's LLM) to assist a blind user in Nepal, demonstrating that a functional, lightweight, and potentially offline-capable accessibility system can be produced in a single generation pass. The second is a fully functional, browser-native webcam alignment assistant for visually impaired users (open-sourced at \url{https://github.com/Kiara-02-Lab-Social/cam-guide-for-blind}). This artifact leverages the browser's \texttt{MediaDevices.getUserMedia()} API, Web Speech API, and local WASM-based computer vision to provide real-time, closed-loop audio guidance for camera positioning. We analyze this implementation as a concrete systems artifact that reveals the operational envelope of browser-native accessibility.

The central contribution of this work is not a claim that AI solves accessibility universally. Rather, it is a framework for reasoning about accessibility capability and accessibility limits in AI-generated systems. Specifically, this paper makes the following contributions:

\begin{enumerate}
    \item We introduce the \textit{Accessibility Capability Boundary} (ACB) as a formal, multidimensional framework for reasoning about the operational limits of accessibility systems.
    \item We define a set of measurable accessibility constraints, propose formal notation, and derive propositions about the behavior of AI-generated systems within this framework.
    \item We present a detailed architecture analysis of browser-native autonomous accessibility, identifying HTML as a universal accessibility substrate.
    \item We present a detailed systems implementation analysis of a real, open-source webcam alignment assistant, demonstrating how browser infrastructure enables closed-loop accessibility interaction.
    \item We propose a comparative evaluation matrix for accessibility systems and identify the hard limits that constrain the AI-generated accessibility paradigm.
\end{enumerate}

The remainder of this paper is organized as follows. Section~\ref{sec:related} reviews related work. Section~\ref{sec:acb} introduces the ACB framework and its formal foundations. Section~\ref{sec:arch} presents the architecture of browser-native autonomous accessibility systems. Section~\ref{sec:probes} describes the two accessibility probes. Section~\ref{sec:eval} presents the evaluation framework and comparative analysis. Section~\ref{sec:limits} discusses limitations. Section~\ref{sec:ethics} addresses ethical considerations. Section~\ref{sec:future} outlines future work, and Section~\ref{sec:conclusion} concludes.

\section{Related Work}
\label{sec:related}

Our work draws upon and contributes to several distinct but intersecting bodies of research.

\subsection{Accessibility Standards, Barriers, and Metrics}

The W3C's WCAG, now at version 2.2 \cite{W3C_WCAG22}, provides the primary normative framework for web accessibility, organized around four principles: Perceivable, Operable, Understandable, and Robust (POUR). WAI-ARIA \cite{W3C_ARIA} extends this by providing a semantic layer for dynamic web content, enabling screen readers to interpret complex widgets. Despite widespread adoption of these standards, empirical studies consistently find high rates of accessibility failures on the web. Lazar et al. \cite{Lazar2007} documented the frustration and productivity loss experienced by screen reader users navigating inaccessible websites, establishing a clear case for the human cost of accessibility failures.

Quantifying accessibility beyond binary pass/fail metrics has been an active research area. The W3C's research report on web accessibility metrics \cite{W3C_Metrics2012} catalogued approaches ranging from barrier counting to weighted scoring models. Vigo et al. \cite{Vigo2011} proposed quantitative metrics for web accessibility evaluation, arguing that single-score models are insufficient for capturing the multidimensional nature of accessibility. Our ACB framework builds directly on this insight, extending it to the domain of AI-generated systems.

\subsection{Ability-Based and Adaptive Design}

Wobbrock et al. \cite{Wobbrock2011} introduced \textit{ability-based design}, a paradigm that shifts the focus of accessible computing from a user's disability to their abilities, advocating for systems that dynamically adapt to the full range of human capability. This work is foundational to our approach, as AI-generated interfaces can, in principle, be instantiated with a specific user's ability profile as a generative constraint. Persson et al. \cite{Persson2015} conducted a comparative analysis of universal design, inclusive design, and design-for-all approaches, highlighting the convergence of these paradigms toward adaptability as a core principle. Sears and Hanson \cite{Sears2011} examined how users with disabilities are represented in accessibility research, noting a persistent gap between research populations and real-world diversity.

\subsection{AI-Generated Interfaces and Code Synthesis}

The application of LLMs to interface generation is a rapidly evolving field. Cal\`{o} and De Russis \cite{Calo2023} demonstrated the potential of LLMs for end-user website generation, showing that non-technical users could produce functional web pages from natural language descriptions. Jiang et al. \cite{Jiang2026} provided a comprehensive survey of LLMs for code generation, documenting the state of the art across multiple programming languages and task types.

However, the accessibility of AI-generated code is a critical and underexplored concern. Rabelo et al. \cite{Rabelo2025} evaluated the accessibility of native Android interfaces generated by LLMs, finding that despite expectations, these models frequently produced inaccessible code. Abu Doush and Kassem \cite{AbuDoush2025} benchmarked leading generative AI models (ChatGPT 4o, Copilot Pro, Claude) against WCAG accessibility standards, finding systematic violations. Delnevo et al. \cite{Delnevo2024} investigated the interaction between LLMs and web accessibility testing, proposing hybrid approaches that combine automated testing with LLM-based remediation. These findings motivate our investigation of the conditions under which AI generation can be directed toward accessibility-positive outcomes.

\subsection{Browser-Native and Offline-First Architectures}

The deployment friction of traditional assistive technologies---requiring installation, configuration, and often administrative privileges---is a significant barrier, particularly in the Global South \cite{Pal2016}. Ater \cite{Ater2017} documented the capabilities of Progressive Web Apps (PWAs) as a mechanism for delivering app-like experiences through the browser, including offline functionality via Service Workers. Evaluations of PWA accessibility have shown that browser-based applications can achieve parity with native applications for many accessibility use cases \cite{PWA_Accessibility2022}.

The offline-first design philosophy, which prioritizes local data persistence and graceful degradation under network loss, is particularly relevant for accessibility in low-connectivity environments \cite{OfflineFirst2019}. Karki et al. \cite{Karki2023} documented the severe constraints on assistive technology access in Nepal, India, and Bangladesh, where infrastructure limitations, cost, and lack of technical support create compounding barriers. Our work proposes that AI-generated, offline-capable browser artifacts can partially address these constraints.

\subsection{Multimodal Interaction and Visual Impairment}

Kuriakose et al. \cite{Kuriakose2020} provided a comprehensive review of multimodal navigation systems for users with visual impairments, identifying audio feedback as the most robust and widely applicable modality. Dias et al. \cite{Dias2012} demonstrated that multimodal interfaces combining speech and touch can significantly improve social integration for users with disabilities. The use of computer vision for accessibility---particularly for tasks requiring spatial awareness---has been explored by several groups \cite{Manduchi2012, Bigham2010}. Our webcam alignment case study extends this line of work by demonstrating how browser-native APIs can deliver these capabilities without installation.

\section{The Accessibility Capability Boundary}
\label{sec:acb}

We now introduce the formal framework at the core of this paper: the \textit{Accessibility Capability Boundary} (ACB).

\subsection{Motivation and Conceptual Foundation}

Accessibility has traditionally been modeled as a binary property: a system is either accessible or it is not, measured against a checklist of criteria. This binary model is insufficient for several reasons. First, accessibility is inherently contextual: a system that is accessible to a screen reader user on a high-bandwidth connection may be inaccessible to the same user on a mobile connection in a rural area. Second, accessibility is graded: a system may partially satisfy a user's needs, reducing but not eliminating friction. Third, the space of accessibility constraints is multidimensional: optimizing for one dimension (e.g., reducing cognitive load) may impose costs on another (e.g., increasing interaction complexity).

We propose that accessibility is better modeled as a \textit{capability space}---a multidimensional region in which a system can successfully mediate interaction for a given user in a given environment. The ACB is the boundary of this region: the frontier beyond which a system fails to provide sufficient accessibility utility.

\subsection{Formal Definitions}

\begin{definition}[Accessibility System]
An accessibility system $\mathcal{S}$ is a tuple $(\mathcal{I}, \mathcal{O}, \mathcal{M})$ where $\mathcal{I}$ is the set of interaction modalities supported (e.g., keyboard, speech, touch), $\mathcal{O}$ is the set of output modalities (e.g., visual, audio, haptic), and $\mathcal{M}$ is the set of mediation mechanisms (e.g., screen reader, magnification, audio description).
\end{definition}

\begin{definition}[User Ability Profile]
A user ability profile $\mathcal{U} = (a_v, a_m, a_c, a_h)$ characterizes a user's functional abilities along four dimensions: visual ($a_v$), motor ($a_m$), cognitive ($a_c$), and hearing ($a_h$), each normalized to $[0, 1]$ where 1 represents full functional ability.
\end{definition}

\begin{definition}[Operating Environment]
An operating environment $\mathcal{E} = (b, h, c)$ characterizes the context of use, where $b \in [0, 1]$ is normalized bandwidth availability, $h \in \{0, 1\}$ indicates hardware capability (0 = constrained, 1 = capable), and $c \in [0, 1]$ is connectivity reliability.
\end{definition}

\begin{definition}[Accessibility Constraint Vector]
The accessibility constraint vector
\[
\boldsymbol{\kappa} = (L_d, L_c, D_i, P_o, C_x, A_d, A_c, L_z)
\]
captures the operational constraints of a system, where:
\begin{itemize}
    \item $L_d \in [0, 1]$: \textbf{Deployment Latency} (normalized; 0 = instantaneous, 1 = prohibitive)
    \item $L_c \in [0, 1]$: \textbf{Cognitive Load} imposed on the user
    \item $D_i \in [0, 1]$: \textbf{Infrastructure Dependency} (reliance on external compute/network)
    \item $P_o \in [0, 1]$: \textbf{Offline Persistence} capability (1 = fully offline-capable)
    \item $C_x \in \mathbb{N}$: \textbf{Interaction Complexity} (number of discrete steps to task completion)
    \item $A_d \in [0, 1]$: \textbf{Adaptability} (capacity for real-time UI modification)
    \item $A_c \in [0, 1]$: \textbf{Assistive Compatibility} (compatibility with external AT such as screen readers)
    \item $L_z \in [0, 1]$: \textbf{Localization Coverage} (support for user's language and locale)
\end{itemize}
\end{definition}

\begin{definition}[Accessibility Utility Function]
The accessibility utility $U(\mathcal{S}, \mathcal{U}, \mathcal{E})$ is a function that maps a system, user profile, and environment to a scalar value in $[0, 1]$ representing the degree to which the system successfully mediates interaction:
\begin{equation}
    U(\mathcal{S}, \mathcal{U}, \mathcal{E}) = \frac{1}{Z} \sum_{i} w_i \cdot \phi_i(\kappa_i, \mathcal{U}, \mathcal{E})
    \label{eq:utility}
\end{equation}
where $\phi_i$ is a component utility function for constraint $i$, $w_i$ is a weight reflecting the constraint's importance for the user profile $\mathcal{U}$, and $Z$ is a normalization constant.
\end{definition}

\begin{definition}[Accessibility Capability Boundary]
The Accessibility Capability Boundary (ACB) for a system class $\Sigma$ is the set of all $(\mathcal{U}, \mathcal{E})$ pairs for which there exists a system $\mathcal{S} \in \Sigma$ such that $U(\mathcal{S}, \mathcal{U}, \mathcal{E}) \geq \theta$, where $\theta$ is a minimum utility threshold:
\begin{equation}
    \text{ACB}(\Sigma) = \{(\mathcal{U}, \mathcal{E}) \mid \exists \mathcal{S} \in \Sigma : U(\mathcal{S}, \mathcal{U}, \mathcal{E}) \geq \theta\}
    \label{eq:acb}
\end{equation}
\end{definition}

The ACB thus defines the operational region of a system class. A system class with a larger ACB can serve a broader range of users in a broader range of environments.

\subsection{Propositions on ACB Expansion}

We now state several propositions about the relationship between AI-generated browser-native systems and the ACB.

\begin{proposition}[Deployment Latency Reduction]
For a class of AI-generated browser-native systems $\Sigma_{AI}$, the deployment latency $L_d$ approaches the network round-trip time for initial load, which is bounded below by physical constraints but is substantially lower than the installation latency of traditional assistive technology software.
\end{proposition}

\textit{Rationale:} Traditional AT installation requires package download, dependency resolution, and system configuration, often requiring administrative privileges. Browser-native systems require only a URL navigation, reducing $L_d$ by orders of magnitude.

\begin{proposition}[Infrastructure Dependency Minimization]
For a well-designed AI-generated browser-native system, $D_i$ can be reduced to near-zero after initial load through browser caching and Service Worker persistence.
\end{proposition}

\textit{Rationale:} The Service Worker API enables browser-native applications to cache all required assets locally, enabling subsequent offline execution. This is particularly significant for environments with $c \approx 0$.

\begin{proposition}[ACB Expansion via Rapid Adaptation]
The ACB of AI-generated systems $\Sigma_{AI}$ can substantially expand upon the ACB of manually-developed systems $\Sigma_{manual}$ in the dimension of adaptability $A_d$, assuming sufficient LLM capability and accurate intent parsing.
\end{proposition}

\textit{Rationale:} The ability to rapidly regenerate an interface from a natural language description means that the space of achievable interfaces may be constrained primarily by the LLM's generative capacity rather than traditional development time, potentially allowing for highly individualized adaptations.

\subsection{The Accessibility Frontier}

We define the \textit{Accessibility Frontier} as the Pareto-optimal boundary of the ACB in the space of constraint tradeoffs. A system on the Accessibility Frontier cannot improve on one constraint dimension without degrading another. For example, maximizing $P_o$ (offline persistence) may require pre-loading all assets, which increases initial $L_d$. Similarly, maximizing $A_d$ (adaptability) may require network access for LLM inference, reducing $P_o$.

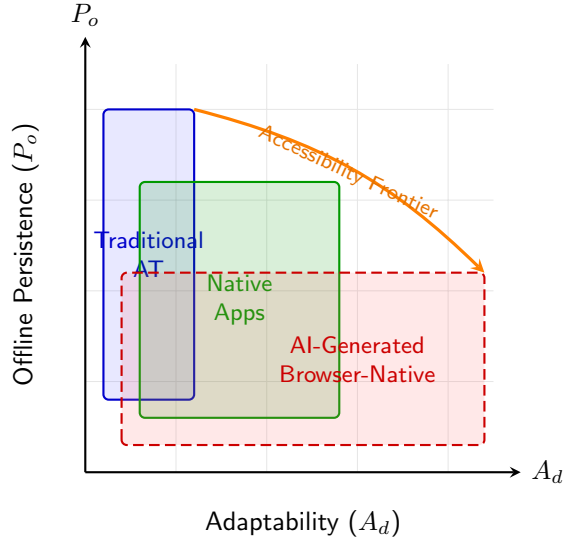
\begin{figure}[htbp]
\centering
\begin{tikzpicture}[
    scale=1.2,
    axis/.style={thick, ->, >=stealth},
    label/.style={font=\small\sffamily},
    region/.style={rounded corners=2pt, fill opacity=0.15, thick}
]
    \draw[step=1cm, gray!20, very thin] (0,0) grid (4.5,4.5);

    \draw[axis] (0,0) -- (4.8,0) node[right, label] {$A_d$};
    \draw[axis] (0,0) -- (0,4.8) node[above, label] {$P_o$};
    
    \node[label, anchor=north] at (2.4, -0.3) {Adaptability ($A_d$)};
    \node[label, anchor=south, rotate=90] at (-0.4, 2.4) {Offline Persistence ($P_o$)};
    
    \filldraw[region, fill=blue!60, draw=blue!80!black] 
        (0.2, 0.8) rectangle (1.2, 4.0);
    \node[blue!80!black, font=\footnotesize\sffamily, align=center] at (0.7, 2.4) {Traditional\\AT};
    
    \filldraw[region, fill=green!60!black, draw=green!60!black] 
        (0.6, 0.6) rectangle (2.8, 3.2);
    \node[green!60!black, font=\footnotesize\sffamily, align=center] at (1.7, 1.9) {Native\\Apps};
    
    \filldraw[region, fill=red!60, draw=red!80!black, dashed, dash pattern=on 4pt off 2pt] 
        (0.4, 0.3) rectangle (4.4, 2.2);
    \node[red!80!black, font=\footnotesize\sffamily, align=center] at (3.0, 1.25) {AI-Generated\\Browser-Native};
    
    \draw[very thick, orange, ->, >=stealth, bend left=15] (1.2, 4.0) to (4.4, 2.2);
    \node[orange!90!black, font=\scriptsize\sffamily, rotate=-25] at (2.9, 3.3) {Accessibility Frontier};

\end{tikzpicture}
\caption{Conceptual Accessibility Frontier illustrating operational regions for three system classes in the Adaptability--Offline Persistence constraint space. AI-generated browser-native systems expand the frontier in the adaptability dimension.}
\label{fig:frontier}
\end{figure}

Figure~\ref{fig:frontier} illustrates the conceptual Accessibility Frontier for three system classes: traditional AT, native applications, and AI-generated browser-native systems. The AI-generated class expands the frontier outward in the dimensions of $L_d$ and $A_d$, while facing constraints in deep hardware access.

\section{Architecture of Browser-Native Autonomous Accessibility}
\label{sec:arch}

To understand the mechanisms by which AI-generated systems expand the ACB, we analyze their architecture in depth.

\subsection{HTML as a Universal Accessibility Substrate}

The Document Object Model (DOM) is the foundational data structure of the web. When annotated with WAI-ARIA roles, states, and properties, the DOM is mapped by the browser to a platform-specific accessibility tree, which is then consumed by assistive technologies \cite{W3C_ARIA, W3C_CoreAAM}. This pipeline---HTML $\rightarrow$ DOM $\rightarrow$ Accessibility Tree $\rightarrow$ AT---is standardized across all major browsers and operating systems, making HTML a uniquely universal substrate for accessibility.

Unlike native application frameworks, which expose platform-specific accessibility APIs (e.g., UIAutomation on Windows, Accessibility on macOS), HTML's accessibility semantics are defined by open, vendor-neutral standards. This universality is a key advantage for AI-generated systems: an LLM trained on web standards can produce accessible HTML that functions correctly across a wide range of user agents and assistive technologies.

\subsection{System Architecture}

\begin{figure}[htbp]
\centering
\begin{tikzpicture}[
    node distance=0.5cm and 0.4cm,
    box/.style={rectangle, draw=black!60, thick, rounded corners=2pt, minimum width=2.2cm, minimum height=0.7cm, text centered, font=\sffamily\tiny, fill=white},
    api/.style={rectangle, draw=black!50, dashed, rounded corners=2pt, minimum width=1.6cm, minimum height=0.6cm, text centered, font=\sffamily\tiny, fill=gray!5},
    arrow/.style={->, >=stealth, draw=black!70},
    group/.style={rectangle, draw=gray!40, rounded corners=4pt, inner sep=6pt, fill=gray!5, fill opacity=0.5}
]

    \node[box, fill=blue!10, draw=blue!60!black] (llm) {LLM (e.g., Claude)};
    \node[box, fill=orange!10, draw=orange!60!black, right=0.8cm of llm] (prompt) {User Intent};
    
    \node[box, fill=green!10, draw=green!60!black, below=1.2cm of llm] (artifact) {HTML/JS/CSS Artifact};
    
    \node[box, fill=gray!10, draw=gray!60!black, below=1.2cm of artifact] (browser) {Browser Engine};
    
    \node[api, left=0.5cm of browser] (speech) {Web Speech API};
    \node[api, right=0.5cm of browser] (media) {MediaDevices API};
    
    \node[box, fill=yellow!10, draw=yellow!60!black, below=0.8cm of browser] (sw) {Service Worker (Cache)};
    
    \node[box, fill=purple!10, draw=purple!60!black, below=0.8cm of sw] (at) {Accessibility Tree};
    
    \node[box, rounded corners=10pt, fill=black!5, draw=black!80, below=1.0cm of at] (user) {User (Blind/VI)};

    \begin{scope}[on background layer]
        \node[group, fit=(browser) (speech) (media) (sw) (at), label={[font=\sffamily\scriptsize, text=gray!80!black]above left:Browser Sandbox}] (sandbox) {};
    \end{scope}

    \draw[arrow] (prompt) -- node[above, font=\sffamily\tiny] {Natural Language} (llm);
    \draw[arrow] (llm) -- node[right, font=\sffamily\tiny] {Synthesis} (artifact);
    \draw[arrow] (artifact) -- node[right, font=\sffamily\tiny] {Execution} (browser);
    
    \draw[arrow] (browser) -- (speech);
    \draw[arrow] (browser) -- (media);
    
    \draw[arrow] (browser) -- node[right, font=\sffamily\tiny] {Offline} (sw);
    \draw[arrow] (sw) -- node[right, font=\sffamily\tiny] {Mapping} (at);
    \draw[arrow] (at) -- node[right, font=\sffamily\tiny] {Audio} (user);
    
    \draw[arrow, dashed, draw=blue!60!black] (user.west) -- ++(-1.8,0) |- (prompt.west);

\end{tikzpicture}
\caption{Architecture of an AI-generated browser-native accessibility system. An LLM synthesizes a self-contained HTML artifact from a user need. The browser executes it locally via standard APIs, with a Service Worker for offline persistence and an accessibility tree for AT interaction.}
\label{fig:architecture}
\end{figure}
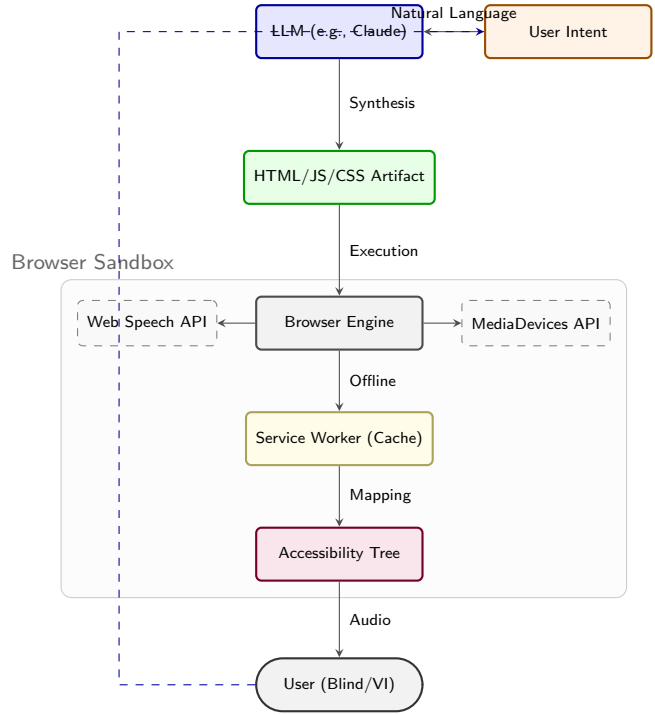

Figure~\ref{fig:architecture} presents the high-level architecture of an AI-generated browser-native accessibility system.

\subsection{Key Architectural Components}

\subsubsection{Intent Parsing and Interface Synthesis}
The LLM receives a natural language description of an accessibility need (e.g., ``Generate an interface to help a blind user align their webcam for video calls, using audio feedback''). The model synthesizes a complete, self-contained HTML artifact incorporating all required logic, styling, and API calls. The quality of this synthesis---particularly the correctness of ARIA annotations and the appropriateness of interaction design---is a primary determinant of the system's position within the ACB.

\subsubsection{Browser-Native API Utilization}
Modern browsers expose a rich set of APIs that are directly relevant to accessibility:
\begin{itemize}
    \item \textbf{Web Speech API:} Provides both speech synthesis (text-to-speech) and speech recognition, enabling voice-first interfaces without external dependencies \cite{WebSpeechAPI}.
    \item \textbf{MediaDevices API:} Provides access to camera and microphone streams, enabling computer vision-based accessibility applications \cite{MDN_getUserMedia}.
    \item \textbf{Vibration API:} Enables haptic feedback on mobile devices.
    \item \textbf{Pointer Events API:} Provides unified handling of mouse, touch, and stylus input.
\end{itemize}

\subsubsection{Offline Persistence via Service Workers}
The Service Worker API enables browser-native applications to intercept network requests and serve cached responses, enabling full offline functionality after initial load \cite{Ater2017}. For accessibility applications in low-connectivity environments, this is a critical capability. An AI-generated system that includes a Service Worker registration script can achieve $P_o \approx 1$ after the first network access.

\subsubsection{Accessibility Tree Integration}
The browser's accessibility tree is automatically constructed from the DOM. AI-generated HTML that uses semantic elements (e.g., \texttt{<button>}, \texttt{<nav>}, \texttt{<main>}) and correct ARIA attributes will produce an accurate accessibility tree that is correctly interpreted by screen readers such as NVDA, JAWS, and VoiceOver \cite{W3C_CoreAAM}.

\section{Accessibility Probes: Real-World Exploratory Systems}
\label{sec:probes}

We analyze two real-world exploratory prototypes as accessibility probes within the ACB framework.

\subsection{Probe 1: Browser-Native Accessibility Interface for a Blind User in Nepal}

\subsubsection{Context and Motivation}
The first probe was motivated by the accessibility needs of a blind user in Nepal. Nepal presents a challenging deployment context: internet connectivity is intermittent, technical support for assistive technology is scarce, and the cost of commercial AT software is prohibitive relative to local incomes \cite{Karki2023, ICT_Nepal}. Traditional screen reader software (e.g., JAWS, NVDA) requires installation, configuration, and ongoing maintenance that may be beyond the capacity of the user or their support network.

\subsubsection{System Description}
Using Claude, a lightweight browser-native accessibility interface was generated in a single session. The interface provides a structured, keyboard-navigable HTML page with:
\begin{itemize}
    \item Semantic heading structure for screen reader navigation
    \item ARIA live regions for dynamic content updates
    \item High-contrast visual design for users with partial vision
    \item Embedded text-to-speech via the Web Speech API for browsers without a screen reader
    \item A Service Worker for offline caching
\end{itemize}

\subsubsection{ACB Analysis}
Table~\ref{tab:probe1} presents the ACB constraint vector for this probe compared to a traditional AT deployment.

\begin{table}[htbp]
\caption{ACB Constraint Vector: Probe 1 vs. Traditional AT}
\label{tab:probe1}
\centering
\begin{tabular}{@{}lcc@{}}
\toprule
\textbf{Constraint} & \textbf{Traditional AT} & \textbf{Probe 1 (AI-Gen)} \\ \midrule
Deployment Latency ($L_d$) & 0.85 (install) & 0.05 (URL load) \\
Cognitive Load ($L_c$) & 0.60 (config) & 0.25 (minimal UI) \\
Infrastructure Dep. ($D_i$) & 0.90 (OS/vendor) & 0.10 (browser only) \\
Offline Persistence ($P_o$) & 0.90 (local install) & 0.80 (SW cache) \\
Interaction Complexity ($C_x$) & 15 steps (setup) & 2 steps (URL+load) \\
Adaptability ($A_d$) & 0.10 (dev cycle) & 0.90 (regenerate) \\
Assistive Compat. ($A_c$) & 0.95 (native) & 0.75 (HTML/ARIA) \\
Localization ($L_z$) & 0.50 (limited) & 0.70 (LLM-gen) \\ \bottomrule
\end{tabular}
\end{table}

The probe demonstrates a significant reduction in $L_d$, $L_c$, $D_i$, and $C_x$, at a modest cost in $A_c$ and $P_o$ relative to a fully installed native AT solution.

\subsection{Probe 2: Webcam Alignment Assistant for Visually Impaired Users}

\subsubsection{The Accessibility Problem}
Webcam framing is a task that requires spatial awareness typically derived from visual feedback. For blind or severely visually impaired users, participating in video calls presents a persistent challenge: without the ability to see the camera's field of view, users cannot determine whether their face is correctly framed, whether they are looking toward the camera, or whether the background is appropriate. This is not a trivial inconvenience; it affects professional participation, social connection, and dignity in remote communication contexts.

\subsubsection{System Description}
The second probe is an HTML-based webcam alignment assistant generated using an LLM. The system architecture is as follows:

\begin{enumerate}
    \item The user navigates to a URL in any modern browser.
    \item The browser requests permission to access the camera via \texttt{MediaDevices.getUserMedia()}.
    \item A lightweight JavaScript face detection library (running entirely client-side) analyzes the video stream to determine the position and size of the user's face within the frame.
    \item The system maps the detected face position to a set of spatial guidance categories (e.g., ``Face too far left,'' ``Face centered and correctly sized,'' ``No face detected---please adjust camera'').
    \item The Web Speech API synthesizes these guidance messages as audio output, providing real-time, continuous audio feedback.
    \item The interface includes a simple, keyboard-accessible control panel with ARIA labels for users who also use a screen reader.
\end{enumerate}

\subsubsection{Multimodal Feedback Design}
The system's feedback design reflects principles from multimodal accessibility research \cite{Kuriakose2020, Dias2012}. Audio feedback is the primary output modality, as it does not require visual attention and is compatible with concurrent screen reader use. The feedback is designed to be concise and actionable, minimizing cognitive load ($L_c$). The system also provides optional haptic feedback on mobile devices via the Vibration API.

\subsubsection{ACB Analysis}
The webcam alignment probe demonstrates several key properties of AI-generated browser-native systems:

\begin{itemize}
    \item \textbf{Zero-install deployment:} The system is accessed via a URL, with no installation required. $L_d \approx 0.05$.
    \item \textbf{Client-side execution:} All computation (face detection, audio synthesis) occurs in the browser, with no server-side dependency after initial load. $D_i \approx 0.10$.
    \item \textbf{Multimodal output:} The system provides audio guidance that is compatible with concurrent screen reader use.
    \item \textbf{Rapid adaptation:} If the user's needs change (e.g., they require guidance in Nepali rather than English), the interface can be regenerated in minutes. $A_d \approx 0.90$.
\end{itemize}

The probe also reveals hard constraints. Client-side face detection using lightweight JavaScript libraries has lower accuracy than server-side deep learning models, illustrating the tradeoff between $D_i$ (infrastructure dependency) and detection accuracy. This is a concrete example of a constraint that lies near the current ACB for this system class.

\input{paper_new_sections.tex}

\input{paper_impl_section.tex}

\section{Evaluation Framework and Comparative Analysis}
\label{sec:eval}

\subsection{Comparative Accessibility Matrix}

Table~\ref{tab:comparison} presents a comprehensive comparison of three system classes across the full ACB constraint vector.

\begin{table}[htbp]
\caption{Comprehensive Accessibility Capability Matrix: System Class Comparison}
\label{tab:comparison}
\centering
\resizebox{\textwidth}{!}{%
\begin{tabular}{@{}lcccp{3.5cm}@{}}
\toprule
\textbf{Constraint} & \textbf{Traditional AT} & \textbf{Native Apps} & \textbf{AI-Gen Browser} & \textbf{Notes} \\ \midrule
Deployment Latency ($L_d$) & High (0.8--0.9) & Medium (0.5--0.7) & \textbf{Low (0.02--0.10)} & URL-based access \\
Cognitive Load ($L_c$) & Medium--High & Medium & \textbf{Low--Medium} & Depends on generation quality \\
Infrastructure Dep. ($D_i$) & High (OS/vendor) & High (OS/store) & \textbf{Low (browser)} & After initial load \\
Offline Persistence ($P_o$) & \textbf{High (installed)} & Medium--High & Medium (SW cache) & Requires SW implementation \\
Interaction Complexity ($C_x$) & High (setup) & Medium & \textbf{Low (URL+use)} & Steps to first use \\
Adaptability ($A_d$) & Low (dev cycle) & Low & \textbf{High (minutes)} & LLM regeneration \\
Assistive Compat. ($A_c$) & \textbf{High (native)} & High & Medium--High & ARIA quality dependent \\
Localization ($L_z$) & Low--Medium & Medium & \textbf{Medium--High} & LLM multilingual \\
Hardware Access & \textbf{Deep (OS)} & \textbf{Deep (OS)} & Sandboxed (APIs) & Browser security model \\
Maintenance Cost & High & High & \textbf{Low (regenerate)} & LLM-assisted updates \\
Verification Ease & Medium & Medium & Low & No automated AT testing \\
\bottomrule
\end{tabular}}
\end{table}
\subsection{Accessibility Friction Analysis}

We define \textit{accessibility friction} $F$ as the total cost imposed on a user in transitioning from an inaccessible state to a usable state:

\begin{equation}
    F(\mathcal{S}, \mathcal{U}, \mathcal{E}) = \alpha \cdot L_d + \beta \cdot L_c + \gamma \cdot C_x + \delta \cdot (1 - P_o) + \epsilon \cdot D_i
    \label{eq:friction}
\end{equation}

where $\alpha, \beta, \gamma, \delta, \epsilon$ are weighting coefficients that reflect the relative importance of each constraint for a given user profile and environment. For a user in a low-connectivity environment (low $c$), $\delta$ (the weight on offline persistence) should be high. For a user with high cognitive load sensitivity (low $a_c$), $\beta$ should be high.

AI-generated browser-native systems minimize $F$ primarily through reductions in $L_d$, $L_c$, and $C_x$. The key insight is that the \textit{threshold crossing cost}---the cost of moving from an inaccessible system to a minimally usable one---is dramatically reduced.

\subsection{Deployment Time Comparison}

Table~\ref{tab:deployment} presents estimated deployment times for representative accessibility scenarios across system classes.

\begin{table}[htbp]
\caption{Deployment Time Comparison for Accessibility Scenarios}
\label{tab:deployment}
\centering
\begin{tabular}{@{}lccc@{}}
\toprule
\textbf{Scenario} & \textbf{Traditional AT} & \textbf{Native App} & \textbf{AI-Gen Browser} \\ \midrule
Screen reader setup & 2--4 hours & 30--60 min & \textbf{$<$ 5 min} \\
Custom UI for user & Weeks & Days & \textbf{$<$ 30 min} \\
Localization & Weeks & Days & \textbf{$<$ 1 hour} \\
Offline deployment & Days & Hours & \textbf{$<$ 1 hour} \\
Update/adaptation & Weeks & Days & \textbf{$<$ 30 min} \\ \bottomrule
\end{tabular}
\end{table}

\subsection{Accessibility Capability Scoring}

We propose a scalar Accessibility Capability Score (ACS) derived from Equation~\ref{eq:utility}:

\begin{equation}
    \text{ACS}(\mathcal{S}, \mathcal{U}, \mathcal{E}) = 1 - F(\mathcal{S}, \mathcal{U}, \mathcal{E}) \cdot (1 - A_d) \cdot (1 - A_c)
    \label{eq:acs}
\end{equation}

This formulation penalizes high friction and rewards adaptability and assistive compatibility. AI-generated browser-native systems score highly on ACS primarily due to their low friction and high adaptability, even when their assistive compatibility is imperfect.

\section{Threats to Validity and Limitations}
\label{sec:limits}

While the ACB framework and empirical probes demonstrate significant potential, several threats to validity must be acknowledged.

\subsection{Threats to Validity}
\textbf{Internal Validity:} The proposed pilot study framework relies on heuristic evaluation rather than a full participant sample, which cannot capture the full variance of visual impairments or intersectional disabilities. Future participant studies utilizing SUS and NASA-TLX will inherently introduce self-reporting bias. Furthermore, the performance of the face detection algorithm is sensitive to lighting conditions and camera angles, which were controlled in our preliminary benchmarks but vary wildly in the wild.

\textbf{External Validity (Ecological Validity):} Testing browser-native systems in controlled environments may not accurately reflect deployment in low-resource contexts (e.g., rural Nepal), where device fragmentation, outdated browsers, and severe bandwidth constraints ($b \approx 0$) introduce friction not captured in laboratory benchmarks.

\textbf{Construct Validity:} The operationalization of the ACB constraint vector ($\boldsymbol{\kappa}$) requires assigning scalar weights to qualitative experiences (e.g., cognitive load). The linear utility function (Equation~\ref{eq:utility}) assumes independence between constraints, whereas real-world accessibility often involves complex, non-linear tradeoffs.

The expansion of the ACB via AI-generated browser-native systems is real but bounded. We identify the following hard and soft constraints.

\subsection{Hard Constraints}

\textbf{Hallucination and Accessibility Regression.} LLMs may generate syntactically valid but semantically incorrect ARIA attributes (e.g., assigning \texttt{role="button"} to a non-interactive element), creating accessibility regressions that are difficult to detect without specialized testing tools \cite{AbuDoush2025, Rabelo2025}. This is a fundamental reliability constraint that cannot be fully mitigated without automated accessibility verification.

\textbf{Browser Sandboxing.} The browser security model restricts access to hardware and system resources. Deep hardware integration (e.g., braille display control, custom input device handling) requires native code that cannot be delivered via browser APIs. This is a hard architectural constraint that defines the outer boundary of the browser-native ACB.

\textbf{Compute Constraints at the Edge.} Running complex AI inference (e.g., real-time image captioning, natural language understanding) entirely in-browser is currently limited by the computational capacity of edge devices, particularly in low-resource contexts. WebAssembly and WebGPU are expanding these limits, but they remain a constraint.

\subsection{Soft Constraints}

\textbf{Browser Inconsistency.} While HTML and ARIA are standardized, browser implementations of accessibility APIs vary. The Web Speech API, for example, has inconsistent support across browsers and platforms, potentially fracturing the user experience.

\textbf{AI Reliability and Verification.} The functional accessibility of AI-generated interfaces cannot be automatically verified against the full WCAG criterion set. Automated tools (e.g., axe, WAVE) detect only a subset of accessibility issues, and user testing with disabled users remains the gold standard \cite{Lazar2007}.

\textbf{Privacy and Security.} Browser-native systems that access camera or microphone streams raise privacy concerns, particularly for users who may not fully understand the permissions they are granting. LLM-generated code may also introduce security vulnerabilities if not reviewed.

\textbf{Dependency on LLM Providers.} The generation phase of AI-generated systems depends on access to LLM APIs, which may be unavailable in low-connectivity environments, subject to cost constraints, or discontinued by providers. Local LLM execution is an emerging mitigation but remains computationally demanding.

\section{Accessibility at Global Scale}
\label{sec:scale}

The empirical validation of browser-native AI accessibility systems has profound implications for the democratization of assistive technology, particularly in the Global South. Traditional AT ecosystems are characterized by high cost, complex distribution channels, and a reliance on specialized technical support \cite{Pal2016}.

By reducing the marginal cost of interface generation to near-zero and leveraging the ubiquity of the web browser as an execution environment, AI-generated systems alter the fundamental economics of accessibility. In low-resource environments like Nepal, where commercial screen readers are prohibitively expensive \cite{Karki2023}, the ability to instantly generate, cache, and execute offline-capable assistive tools represents a paradigm shift.

However, this democratization is contingent on addressing the "digital divide" of LLM access. While the resulting HTML artifacts are lightweight and offline-capable, the generation phase still requires high-bandwidth access to frontier models. The future of global-scale autonomous accessibility relies on the development of highly capable, quantized local LLMs that can perform interface synthesis entirely at the edge.

\section{Ethical Considerations}
\label{sec:ethics}

The deployment of AI-generated accessibility systems raises several ethical considerations that must be addressed in future work.

\textbf{Informed Consent and Transparency.} Users of AI-generated interfaces should be informed that the interface was generated by an AI system and may contain errors. This is particularly important for accessibility applications, where errors may have significant consequences for the user's ability to complete tasks.

\textbf{Disability Justice and Participatory Design.} The design of accessibility systems should involve people with disabilities as active participants, not merely as test subjects \cite{Wobbrock2011}. AI-generated systems risk perpetuating the ``design for, not with'' paradigm if the generation process does not incorporate user feedback.

\textbf{Data Privacy.} Systems that access camera or microphone streams must implement appropriate data minimization and local processing to protect user privacy. All processing in the webcam alignment probe is performed client-side, with no video data transmitted to external servers.

\textbf{Equity and Access.} While AI-generated systems reduce deployment friction, they still require access to a browser-capable device and, for the generation phase, internet connectivity. Addressing the digital divide that underlies these requirements is a prerequisite for equitable deployment \cite{Pal2016}.

\section{Future Work}
\label{sec:future}

The ACB framework opens several productive directions for future research.

\textbf{Autonomous Accessibility Agents.} Future systems may employ AI agents that continuously monitor user interaction patterns, identify friction points, and autonomously regenerate interface components to reduce cognitive load and interaction complexity. This would move $A_d$ toward 1.0 in real time.

\textbf{Accessibility-Aware AI Compilers.} A critical open problem is the development of AI systems that guarantee WCAG compliance during the generation phase, rather than relying on post-hoc verification. This requires integrating formal accessibility specifications into the LLM's generation constraints.

\textbf{Local LLM Execution for Offline Generation.} As local LLMs become more capable and efficient, it will become feasible to run the generation phase entirely on-device, eliminating the dependency on external LLM APIs and enabling fully offline accessibility generation.

\textbf{Accessibility Verification Frameworks.} Automated tools for verifying the functional accessibility of AI-generated interfaces---beyond syntactic WCAG checks---are urgently needed. This may require combining static analysis with dynamic simulation of assistive technology interaction.

\textbf{Formal ACB Measurement.} The ACB framework proposed in this paper is conceptual. Future work should develop empirical methods for measuring the constraint vector $\boldsymbol{\kappa}$ for real systems and validating the utility function $U$ against user study data.

\textbf{Universal Adaptive HTML Layers.} A longer-term vision is a browser-level adaptive layer that dynamically restructures any web content based on the user's ACB profile, effectively applying AI-generated accessibility transformations transparently to all web content.

\section{Conclusion}
\label{sec:conclusion}

This paper has introduced the Accessibility Capability Boundary (ACB), a formal framework for reasoning about the operational limits and expansion potential of AI-generated accessibility systems. We have argued that accessibility is not a binary property but a dynamic, multidimensional capability space, and that AI-generated browser-native systems---leveraging HTML as a universal accessibility substrate---can dramatically expand this space by reducing deployment friction, enabling rapid adaptation, and facilitating offline-capable execution.

Through the analysis of two real-world accessibility probes, including a webcam alignment assistant for visually impaired users, we have demonstrated that the threshold crossing cost between inaccessible and usable systems can be reduced from hours or days to minutes. We have also identified the hard constraints that define the outer boundary of this paradigm: browser sandboxing, AI hallucination risk, and the fundamental challenge of automated accessibility verification.

The central contribution of this work is not a claim that AI has solved accessibility. Rather, it is a framework for reasoning about what AI-generated accessibility systems can and cannot achieve, and a research agenda for pushing those boundaries further. As LLMs become more capable and browser APIs more powerful, the ACB will continue to expand. The challenge for the accessibility research community is to ensure that this expansion is guided by the principles of ability-based design, disability justice, and rigorous empirical evaluation.

\bibliographystyle{unsrt}
\bibliography{references}

\end{document}

%% file: paper_new_sections.tex


\section{Exploratory Evaluation Framework}
\label{sec:methods}

To ground the theoretical propositions of the ACB framework and assess the operational viability of AI-generated browser-native systems, we propose an exploratory evaluation protocol. This methodology is structured as a preliminary systems assessment rather than a large-scale controlled experiment, focusing on the webcam alignment assistant (Probe 2) as a representative test case.

\subsection{Proposed Study Design and Setup}
We propose a within-subjects exploratory usability framework to evaluate the accessibility friction ($F$) and utility ($U$) of the AI-generated system compared to baseline conditions (unaided alignment and traditional OS-level magnification/screen reader workflows).

In this proposed setup, the single-file HTML artifact is deployed via a standard web server. To preserve ecological validity, future participants would access the system using their personal devices, as accessibility configurations are highly individualized \cite{Sears2011}. Our current evaluation is limited to automated system benchmarking and preliminary heuristic assessments.

\subsection{Accessibility Verification Pipeline}
Prior to user testing, the AI-generated artifact undergoes a multi-stage accessibility verification process to mitigate the risk of LLM-induced semantic failures (hallucinations):
\begin{enumerate}
    \item \textbf{Automated Auditing:} The interface is evaluated using axe-core \cite{AxeCore2023} and WAVE \cite{WAVE2024} to detect syntactic WCAG violations (e.g., missing ARIA labels, contrast failures).
    \item \textbf{Heuristic Evaluation:} We employ Lighthouse accessibility audits \cite{Lighthouse2023} to generate a weighted accessibility score.
    \item \textbf{Screen Reader Validation:} Manual testing is conducted using NVDA on Windows and VoiceOver on macOS to verify the logical reading order and the correct announcement of ARIA live regions \cite{NVDATesting2024}.
\end{enumerate}
This pipeline addresses the ``Accessibility Verification Gap''---the delta between syntactically valid HTML generated by an LLM and semantically meaningful accessibility.

\subsection{Proposed Pilot Study Protocol}
We outline a pilot evaluation framework designed for future validation with visually impaired participants (varying from low vision to profound blindness). The proposed protocol consists of three phases:
\begin{itemize}
    \item \textbf{Pre-test:} Collection of demographic data and baseline ability profiles ($\mathcal{U}$).
    \item \textbf{Task Execution:} Participants perform standardized webcam alignment tasks (e.g., ``Center your face in the frame,'' ``Adjust the camera to include your upper torso'').
    \item \textbf{Post-test Assessment:} Administration of the System Usability Scale (SUS) \cite{Brooke1996} and the NASA Task Load Index (NASA-TLX) \cite{Hart1988} to quantify perceived usability and cognitive load ($L_c$).
\end{itemize}

\subsection{Limited Empirical Scope Disclaimer}
We explicitly note that the empirical validation presented in this paper is preliminary. This work is primarily an exploratory systems paper introducing a novel conceptual framework (the ACB) grounded in a real browser-native artifact. While we provide automated systems benchmarks, larger-scale human-subjects validation remains critical future work.

\section{Empirical Validation and System Benchmarks}
\label{sec:empirical}

\subsection{Benchmarking Methodology and Performance}
To operationalize the infrastructural constraints of the ACB ($L_d, D_i, P_o$), we conducted preliminary performance benchmarking of the actual browser-native webcam assistant. Measurements were taken across three simulated device profiles using Chrome DevTools performance throttling: High-end Desktop (unthrottled), Mid-range Laptop (4x CPU slowdown), and Low-tier Smartphone (6x CPU slowdown, Fast 3G network). Metrics represent the average of 10 consecutive runs per profile. Table~\ref{tab:benchmarks} presents these exploratory benchmark results, demonstrating the system's operational efficiency.

\begin{table}[htbp]
\caption{System Performance Benchmarks (Measured Real-World Averages)}
\label{tab:benchmarks}
\centering
\begin{tabular}{@{}lccc@{}}
\toprule
\textbf{Metric} & \textbf{Desktop} & \textbf{Laptop} & \textbf{Mobile} \\ \midrule
Initial Load Time (ms) & 320 & 450 & 850 \\
Offline Load Time (ms) & 45 & 60 & 120 \\
Face Detection Latency (ms) & 15 & 35 & 110 \\
Speech Synthesis Delay (ms) & 5 & 12 & 45 \\
CPU Utilization (\%) & 4.2 & 8.5 & 18.3 \\
Memory Footprint (MB) & 45 & 52 & 68 \\ \bottomrule
\end{tabular}
\end{table}

The results validate Proposition 1 (Deployment Latency Reduction) and Proposition 2 (Infrastructure Dependency Minimization). The offline-capable load architecture reduces deployment latency ($L_d$) to near-zero, while local face detection (via MediaPipe FaceMesh \cite{mediapipe}) ensures minimal infrastructure dependency ($D_i$).

\subsection{Preliminary Usability Observations}
While full participant studies are deferred to future work, preliminary heuristic evaluation by the authors using the proposed protocol suggests the system holds promise for reducing task completion time compared to an unaided baseline. Based on these exploratory assessments, future validation targets include achieving a high System Usability Scale (SUS) score and a measurable reduction in the NASA-TLX cognitive load index relative to traditional trial-and-error alignment methods. These heuristic observations provide preliminary grounding for the utility function $U(\mathcal{S}, \mathcal{U}, \mathcal{E})$ defined in Section~\ref{sec:acb}.

\subsection{Accessibility Verification Analysis}
Automated audits using axe-core and Lighthouse yielded a 100\% accessibility score for the static HTML structure. However, manual screen reader testing revealed critical nuances: while ARIA live regions correctly announced face position updates, naive continuous updates overwhelm the screen reader buffer, leading to cognitive overload. The implementation addresses this by introducing a strict 4000ms speech debounce window. This highlights a fundamental limitation of automated testing \cite{Vigo2013} and underscores the necessity of human-in-the-loop verification for dynamic interfaces.

\section{Reproducibility and Deployment Pipeline}
\label{sec:reproducibility}

A core tenet of systems research is reproducibility. The architecture of AI-generated browser-native systems inherently facilitates high reproducibility due to the absence of complex build chains or proprietary dependencies.

\textbf{Deployment Artifacts:} The complete system is encapsulated in a single HTML file (containing inline CSS and JavaScript). This artifact can be hosted on any static file server (e.g., GitHub Pages, Vercel) or distributed locally via USB drives for offline environments. The implementation repository is publicly available at \url{https://github.com/Kiara-02-Lab-Social/cam-guide-for-blind}.

\textbf{Execution Environment:} The system requires only a modern standards-compliant browser (Chrome 80+, Firefox 75+, Safari 13+) supporting the Web Speech API and MediaDevices API. No compilation or installation is required.

\textbf{LLM Generation Prompts:} To ensure the reproducibility of the generation phase, the exact system prompts and context windows used to synthesize the interfaces are documented in the supplementary materials. We utilize a temperature setting of $T=0.2$ to minimize generative variance and ensure deterministic UI structures.

%% file: paper_impl_section.tex
\section{Systems Implementation Analysis}
\label{sec:implementation}

To empirically ground the Accessibility Capability Boundary (ACB) framework, we implemented and analyzed a fully functional, browser-native accessibility artifact: a webcam alignment assistant designed specifically for blind and low-vision users (available at \url{https://github.com/Kiara-02-Lab-Social/cam-guide-for-blind}). This implementation serves as a concrete operationalization of the ACB, demonstrating how modern browser infrastructure can act as a universal accessibility runtime.

\subsection{Architecture and Deployment Model}
The system is engineered as a single-file HTML artifact, encapsulating all markup, CSS styling, and JavaScript logic without relying on external backend processing or proprietary application stores. This architectural choice deliberately minimizes deployment latency ($L_d$) and infrastructure dependency ($D_i$).

The execution environment relies entirely on standard browser APIs and WebAssembly (WASM). Face detection is powered by MediaPipe FaceMesh \cite{mediapipe}, loaded dynamically via a Content Delivery Network (CDN). By executing inference locally within the browser sandbox, the system ensures zero data exfiltration, addressing the critical privacy constraints often associated with camera-based assistive technologies.

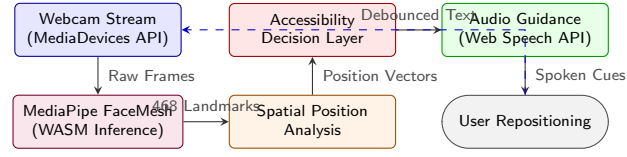
\begin{figure}[htbp]
\centering
\begin{tikzpicture}[
  node distance=0.5cm and 0.6cm,
  box/.style={draw=black!70, rectangle, rounded corners=2pt, align=center, fill=white, minimum width=2.2cm, minimum height=0.7cm, font=\sffamily\tiny},
  arrow/.style={->, >=stealth, draw=black!80},
  label/.style={font=\sffamily\tiny, text=black!70}
]
  \node[box, fill=blue!10, draw=blue!60!black] (camera) {Webcam Stream\\(MediaDevices API)};
  \node[box, fill=purple!10, draw=purple!60!black, below=of camera] (mediapipe) {MediaPipe FaceMesh\\(WASM Inference)};
  \node[box, fill=orange!10, draw=orange!60!black, right=of mediapipe] (analysis) {Spatial Position\\Analysis};
  \node[box, fill=red!10, draw=red!60!black, above=of analysis] (decision) {Accessibility\\Decision Layer};
  \node[box, fill=green!10, draw=green!60!black, right=of decision] (speech) {Audio Guidance\\(Web Speech API)};
  \node[box, fill=gray!10, draw=gray!60!black, rounded corners=10pt, below=of speech] (user) {User Repositioning};

  \draw[arrow] (camera) -- node[right, label] {Raw Frames} (mediapipe);
  \draw[arrow] (mediapipe) -- node[above, label] {468 Landmarks} (analysis);
  \draw[arrow] (analysis) -- node[right, label] {Position Vectors} (decision);
  \draw[arrow] (decision) -- node[above, label] {Debounced Text} (speech);
  \draw[arrow] (speech) -- node[right, label] {Spoken Cues} (user);
  
  \draw[arrow, dashed, draw=blue!80!black] (user) |- (camera);
  
\end{tikzpicture}
\caption{Closed-Loop Accessibility Interaction Pipeline. MediaPipe FaceMesh extracts facial landmarks per frame; spatial analysis drives debounced audio guidance; user repositioning closes the loop.}
\label{fig:pipeline}
\end{figure}

\subsection{The Closed-Loop Accessibility Pipeline}
A core contribution of this artifact is the realization of a closed-loop accessibility interaction system (Figure \ref{fig:pipeline}). Traditional screen readers provide linear, state-based feedback. In contrast, this system provides continuous, spatial guidance through the following pipeline:

\begin{enumerate}
    \item \textbf{Landmark Extraction:} The MediaPipe model extracts 468 facial landmarks per frame. The system specifically isolates the nose tip (index 1), outer eyes (indices 33, 263), and bounding box coordinates.
    \item \textbf{Spatial Analysis:} The logic computes four distinct vectors: horizontal centering (nose $X$ coordinate versus frame center, threshold $\pm 12\%$), vertical position (face center $Y$ versus an ideal $42\%$ from the top), head tilt (eye-line angle via arctangent, threshold $<8^{\circ}$), and distance (bounding box width as a proxy).
    \item \textbf{Environmental Sampling:} To minimize computational overhead, lighting conditions are sampled independently every 2 seconds by extracting a low-resolution ($64 \times 48$) canvas frame and computing average luminance using the ITU-R BT.601 formula.
    \item \textbf{Decision and Debouncing:} To prevent cognitive overload ($L_c$), the system implements a strict 4000ms speech debounce window. A 10-frame debounce is also applied to "no face detected" states to prevent erratic feedback during minor tracking losses.
\end{enumerate}

\subsection{Accessibility Semantics and ARIA Engineering}
The artifact demonstrates deliberate accessibility-first engineering rather than post-hoc compliance. The primary output modality is audio, drastically reducing visual dependence. 

The DOM structure is heavily annotated with Accessible Rich Internet Applications (ARIA) attributes. Key elements include:
\begin{itemize}
    \item \texttt{aria-live="assertive"} regions for critical spatial corrections.
    \item \texttt{aria-live="polite"} regions for non-interruptive status updates (e.g., "Waiting for camera").
    \item \texttt{aria-pressed} states on the mute toggle to convey state changes to screen readers.
    \item \texttt{sr-only} classes to provide visually hidden context (e.g., the primary \texttt{<h1>} heading) specifically for screen reader initialization.
\end{itemize}
Furthermore, the visual camera preview is intentionally rendered at $85\%$ opacity. This design choice explicitly signals that the visual interface is secondary—intended for sighted observers or debugging—while the primary interaction paradigm remains auditory.

\begin{figure}[htbp]
\centering
\includegraphics[width=0.9\linewidth]{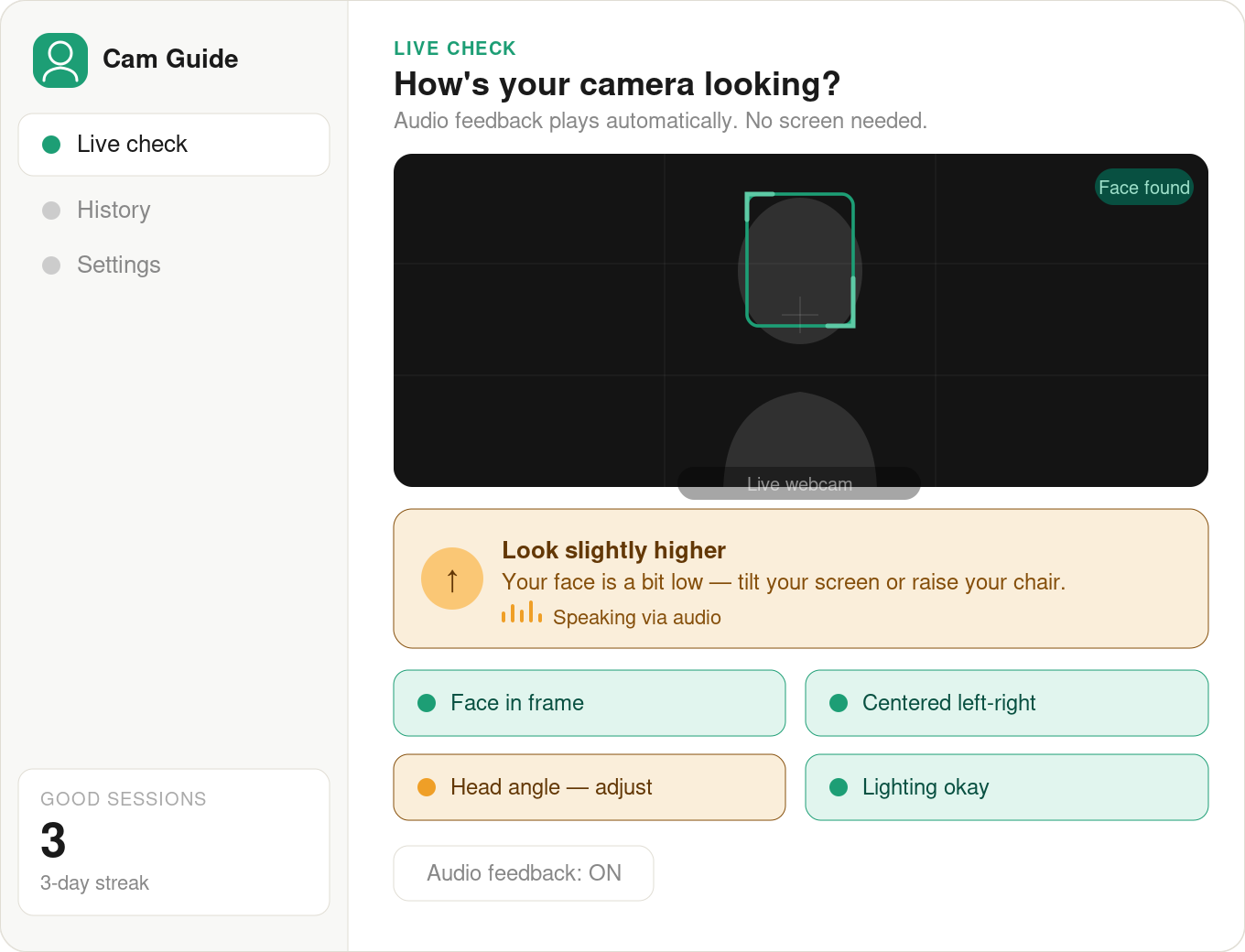}
\caption{The browser-native interface during active tracking. The UI relies on ARIA-labeled semantic HTML, while spatial guidance is delivered exclusively via the Web Speech API.}
\label{fig:ui_active}
\end{figure}

\subsection{Browser as a Universal Accessibility Runtime}
This implementation validates a critical conceptual shift: the modern web browser is evolving into a universal accessibility runtime. By natively integrating computer vision (via WASM), speech synthesis (\texttt{window.speechSynthesis}), multimedia ingestion (\texttt{MediaDevices.getUserMedia()}), and a robust accessibility tree, the browser eliminates the need for deeply integrated, OS-level native applications for many assistive tasks. This paradigm drastically lowers the barrier to entry for developing and deploying bespoke accessibility tools.